\documentclass[12pt]{article}
\usepackage{epsfig,amssymb,amsmath}
\usepackage{jheppub}
 
\usepackage{esint} 
\usepackage{breqn}

\usepackage{soul}

\def \coupling{a}

\def \tr {\mathop{\rm tr}\nolimits}

\def \e  {\mathop{\rm e}\nolimits}

\newcommand\lr[1]{{\left({#1}\right)}}

\newcommand \vev [1] {\langle{#1}\rangle}

\newcommand\re[1]{(\ref{#1})}
\def \qqquad {\qquad\quad}
\def \qqqquad {\qquad\qquad}

\newcommand{\ft}[2]{{\textstyle\frac{#1}{#2}}}

\def\numberbysection{\@addtoreset{equation}{section}
                     \def\theequation{\thesection.\arabic{equation}}}

\preprint{\small  \parbox[t]{25mm}{IPhT-T16/092}}

\title{\Large On instanton effects in the operator product expansion}

\author[a]{Luis F. Alday}
\author[b]{and Gregory P. Korchemsky}

\affiliation[a]{Mathematical Institute, University of Oxford,  Andrew Wiles Building, Radcliffe Observatory Quarter, Woodstock Road, Oxford, OX2 6GG, UK}

\affiliation[b]{Institut de Physique Th\'eorique\footnote{Unit\'e Mixte de Recherche 3681 du CNRS}, Universit\'e Paris Saclay, CNRS, CEA, F-91191 Gif-sur-Yvette}

\abstract{We revisit the computation of instanton effects to various correlation functions in ${\cal N}=4$ SYM
and clarify a controversy existing in the literature regarding their consistency with the OPE and conformal symmetry. 
To check these properties, we examine the conformal partial wave decomposition of four-point correlators involving combinations of half-BPS and Konishi operators and isolate the contribution from the conformal primary scalar operators of twist four.
We demonstrate that the leading instanton correction to this contribution is indeed consistent 
with conformal symmetry and compute the corresponding corrections to the OPE coefficients and the scaling dimensions of such twist-four operators.
Our analysis justifies the regularization procedure used to compute ultraviolet divergent instanton contribution to 
correlation functions involving unprotected operators.
}

\begin{document}

\maketitle

\section{Introduction} 

Four-point correlation functions of half-BPS operators are important quantities  in maximally supersymmetric $\mathcal N=4$ Yang-Mills theory.
They encode the nontrivial dynamics of the theory and have been intensively studied in the past in connection to the AdS/CFT duality. Correlation functions receive quantum corrections which can be separated at weak coupling into perturbative and non-perturbative (instanton) ones.
The latter corrections are exponentially suppressed in the planar limit but they are expected to play a crucial role in restoring $S-$duality.

In $\mathcal N=4$ SYM with the $SU(N)$ gauge group, this symmetry implies invariance under $SL(2,\mathbb Z)$ modular transformations acting on the complexified coupling constant~\cite{Montonen:1977sn,Sen:1994yi,Witten:1995zh}
\begin{equation}
\tau = \frac{\theta}{2\pi} + \frac{4\pi i}{g^2}\, .
\end{equation}  
One of the consequences of $S-$duality is that the above mentioned correlation functions should, in principle, depend on the $\theta-$angle, through 
non-perturbative instanton corrections. Understanding the modular properties of correlation functions requires taking into account instanton effects.
This problem still awaits its solution.
 
The leading instanton corrections to correlation functions can be computed semiclassically by replacing all fields by their classical expressions on the 
instanton background and neglecting quantum fluctuations. In this approximation, the correlation functions are given by finite-dimensional integrals over 
the collective coordinates of the instantons, see {\it e.g.} \cite{Bianchi:1998nk,Belitsky:2000ws}
\begin{equation}
\label{instsc}
\langle O(1) \dots O(n) \rangle_{\rm inst} = \int d\mu_{\rm phys} \e^{-S_{\rm inst}} O(1) \dots O(n)\,, 
\end{equation}
where all operators on the right-hand side are evaluated at the instanton field configuration. In the simplest case of $SU(2)$ gauge group the integration measure for the one-instanton sector takes the form
\begin{equation}\label{measure}
 \int d\mu_{\rm phys} \e^{-S_{\rm inst}} =\frac{g^8}{2^{34}\pi^{10}} \int d^4 x_0 \int_0^\infty \frac{d\rho}{\rho} \int d^8 \xi \int d^8 \bar \eta\,,
\end{equation}
where the bosonic collective coordinates $(\rho,x_0)$ parametrize the size and location of the instanton and $16$ fermion coordinates 
$\xi_\alpha^A$ and $\bar \eta_{\dot \alpha}^A$ (with $\alpha,\dot \alpha=1,2$ and $A=1,\dots,4$) arise due to the $\mathcal N=4$ superconformal
symmetry. For the integral \re{instsc} to be different from zero, the product of operators $O(1) \dots O(n)$ should soak up all $16$ fermion modes.
The corresponding correlations functions are called minimal. In this case, it is possible to generalize \re{instsc} to the $SU(N)$ gauge group and,
in addition, take into account the contribution of an arbitrary number of instantons at large $N$ \cite{Dorey:1999pd}.
   
In this paper we focus on instanton effects in four-point correlation functions of half-BPS operators $O_{\bf 20'}(x,Y)$ made out of scalar fields
\begin{align}
\label{bil}
O_{\bf 20'}(x,Y) &= \frac{1}{g^2} Y_{AB} Y_{CD} \tr(\phi^{AB}\phi^{CD})\,, 
\end{align}
where $Y_{AB}$ is an antisymmetric tensor satisfying $\epsilon^{ABCD} Y_{AB} Y_{CD}=0$. The operator $O_{\bf 20'}(x,Y)$ belongs to the $\bf 20'$ representation of the $SU(4)$ $R-$symmetry group and its scaling dimension is protected from quantum corrections.  Having computed
$\vev{O_{\bf 20'}(1)\dots O_{\bf 20'}(4)}$, we can apply the OPE and decompose it over conformal partial waves corresponding to various
conformal primary operators with $R-$charges in the tensor product $\bf 20'\times \bf 20'$. In what follows we shall restrict our consideration to
conformal operators in the singlet representation of $SU(4)$ with low scaling dimension. They include the Konishi operator with bare
dimension $2$
\begin{align} \label{Ko}
K(x) &= \frac{1}{g^2}\tr(\bar \phi_{AB} \phi^{AB}), 
\end{align}
and four quadrilinear operators with bare dimension $4$ 
\begin{align}
\label{quadrilinear}
&{\cal A}_1= \frac{1}{g^4} \tr(\bar \phi_{AB} \phi^{CD})  \tr(\bar \phi_{CD} \phi^{AB}), ~~~~~{\cal A}_2=\frac{1}{g^4} \tr(\bar \phi_{AB} \phi^{AB})  \tr(\bar \phi_{CD} \phi^{CD}), \cr
&{\cal A}_3= \frac{1}{g^4}\tr(\bar \phi_{AB} \phi^{CD} \bar \phi_{CD} \phi^{AB}),~~~~~~~~~~{\cal A}_4= \frac{1}{g^4}\tr(\bar \phi_{AB} \phi^{AB} \bar \phi_{CD} \phi^{CD}), 
\end{align}
where $\bar \phi_{AB}= \frac{1}{2}\epsilon_{ABCD} \phi^{CD}$. At quantum level, the operators ${\cal A}_i$ mix with each other and
the conformal operators are given by specific linear combinations of those.

The definition of the operators \re{bil}, \re{Ko} and \re{quadrilinear} involves additional powers of the inverse coupling constant, one per 
each scalar field. This reflects our choice for the Lagrangian of $\mathcal N=4$ SYM. Computing instanton corrections it proves convenient to choose 
it in the form $L=1/g^2\tr(-\ft12 F_{\mu\nu}^2  - \ft12 D^\mu \phi^{AB} D_\mu \bar\phi_{AB} + \dots)$. In this case, the corresponding equations of motion 
are coupling independent but free scalar propagator contains an additional factor of $g^2$ (see \re{free}). The operators  \re{bil}, \re{Ko} and \re{quadrilinear} are defined in such a way that their correlation functions do not depend on the coupling constant in the Born approximation.

By virtue of conformal symmetry, the contribution of the operators \re{Ko} and \re{quadrilinear} to the four-point correlation function $\vev{O_{\bf 20'}(1)\dots O_{\bf 20'}(4)}_{\rm inst}$ can be expressed in terms of the scaling dimensions and OPE coefficients defined by the following two- and three-point functions
\begin{align}\label{func}
\langle O_{\bf 20'} O_{\bf 20'} K \rangle_{\rm inst}\,, \qqquad \langle KK \rangle_{\rm inst}\,, \qqquad
\langle O_{\bf 20'} O_{\bf 20'} {\cal A}_i \rangle_{\rm inst}\,, \qqquad \langle {\cal A}_i{\cal A}_j \rangle_{\rm inst}\,.
\end{align}
In the semiclassical approximation, these correlation functions can be computed using \re{instsc}. Since the operators \re{bil}, \re{Ko} and \re{quadrilinear}
are built from scalar fields, we only need the expression for the scalar field on the instanton background in ${\cal N}=4$ SYM. It takes the following general form 
for one-instanton solution
\begin{equation}\label{pro}
\phi^{AB} = \phi^{AB,(2)}+ \phi^{AB,(6)} + \dots\,,
\end{equation}
where $\phi^{AB,(n)}$ denotes the contribution containing $n$ fermion modes. The leading term has been worked out in \cite{Belitsky:2000ws}, while the subleading term has been worked out only recently \cite{Alday:2016tll,Alday:2016jeo}. Notice that $\phi^{AB}$ does not depend on the coupling constant
due to our choice of the Lagrangian. 

Replacing the scalar fields in  \re{bil}, \re{Ko} and \re{quadrilinear} with \re{pro} we find the instanton profile of 
the operators
\begin{equation}\label{K-exp}
O_{\bf 20'}= {1\over g^2} O_{\bf 20'}^{(4)} \,, \qqquad K={1\over g^2} K^{(8)}+\cdots\,, \qqquad{\cal A}_i = {1\over g^4}{\cal A}_i^{(8)} + \cdots\,,
\end{equation} 
where for the Konishi operator the expansion starts with $8$ modes due to vanishing of the leading term, $K^{(4)}=0$. Substituting these
relations into \re{func} and applying \re{instsc} we find that the correlation functions \re{func} are different from zero but have different dependence
on the coupling constant 
\begin{align}\notag
& \langle O_{\bf 20'} O_{\bf 20'} K \rangle_{\rm inst}=O(g^2 q)\,, && \langle KK \rangle_{\rm inst}=O(g^4 q)\,, 
\\[2mm]
& \langle O_{\bf 20'} O_{\bf 20'} {\cal A}_i \rangle_{\rm inst}=O(q)\,, && \langle {\cal A}_i{\cal A}_j \rangle_{\rm inst}=O(q)\,,
\end{align}
where $q=\exp(2\pi i\tau)$ comes from $\e^{-S_{\rm inst}}$ evaluated at the one-instanton configuration. The fact that the expressions in the first line
are suppressed by a power of the coupling constant compared to those in the second line, implies that the contribution from the Konishi operator to  $\langle O_{\bf 20'}O_{\bf 20'}O_{\bf 20'}O_{\bf 20'} \rangle_{\rm inst}$ is subleading. The leading $O(q)$ instanton contribution only comes from the 
 quadrilinear operators \re{quadrilinear}. Note that expansions (\ref{K-exp}) also imply that the instanton contribution to the four point correlators involving the Konishi operator, namely $\langle O_{\bf 20'}O_{\bf 20'}KK \rangle_{\rm inst}$ and $\langle KKKK \rangle_{\rm inst}$, vanish at leading $O(q)$ order. 

The leading instanton contribution to the correlation function of four half-BPS operators was first computed in \cite{Bianchi:1998nk} and its OPE 
decomposition was further analysed in \cite{Arutyunov:2000im}. Later on, the mixing matrix  $\langle {\cal A}_i{\cal A}_j \rangle_{\rm inst}$ was computed in \cite{Kovacs:2003rt} and found  to be
in conflict with the OPE analysis performed in  \cite{Arutyunov:2000im}. A possible reason for such a disagreement could be the fact that instanton corrections to correlators involving unprotected operators involve ultraviolet divergent integrals that need to be regularized,  {\it e.g.} by going slightly 
away from four dimensions in the integral over the position of the instanton in \re{measure}. Although the contribution of the quadrilinear operators 
to $\langle O_{\bf 20'}O_{\bf 20'}O_{\bf 20'}O_{\bf 20'} \rangle_{\rm inst}$ should be finite and  regularization scheme independent, it is not clear a priori that this procedure does not introduce any subtleties. 

The main aim of the present paper is to resolve this puzzle. We do so by revisiting the computation of $\langle {\cal A}_i{\cal A}_j \rangle_{\rm inst}$ and going through a careful OPE analysis. The instanton corrections afffect the mixing matrix of the quadrilinear operators and modify the form
of the conformal primary operators. We compute the leading instanton contribution to scaling dimensions of these operators and their OPE coefficients in the product of operators $O_{\bf 20'}(1)O_{\bf 20'}(2)$ and $K(1) K(2)$. We show that our results are fully consistent with the known expressions for all relevant four point correlators $\langle O_{\bf 20'}O_{\bf 20'}O_{\bf 20'}O_{\bf 20'} \rangle_{\rm inst}$, 
$\langle O_{\bf 20'}O_{\bf 20'}KK \rangle_{\rm inst}$ and $\langle KKKK \rangle_{\rm inst}$.
This not only solves the puzzle mentioned above, but also justifies the regularization procedure that we employed to compute instanton corrections to correlators involving unprotected operators. Furthermore, as a byproduct of our analysis, we determine the leading instanton contribution to the scaling dimension of twist-four operators.

This paper is organised as follows. In section \ref{OPEanal} we review known results regarding instanton corrections to four-point 
correlation functions of half-BPS operators. In section \ref{instresults} we compute  the leading instanton corrections to correlators involving the quadrilinear operators \re{quadrilinear}. In section \ref{comparison} we show that the results obtained in section \ref{instresults} are fully consistent with the OPE decomposition of four-point correlation functions. Section \ref{concl} contains concluding remarks.
In addition, in appendix \ref{appA} we include formulae for various correlations functions in the Born approximation. In appendix \ref{appB} we discuss regularization of divergent integrals arising in the computation of the instanton corrections. We show in appendix \ref{appC} that such integrals, after non-trivial cancelations, lead to finite OPE coefficients.

\section{Four-point correlation functions}
\label{OPEanal}
In this section we review known results regarding instanton corrections to four-point correlators. Furthermore, we perform an OPE analysis focusing on the contribution from the twist four operators mentioned in the introduction. 

We start by considering the four-point correlator of half-BPS operators 
\begin{align}\label{G4}
G_4=\langle O_{\bf 20'}(x_1,Y_1)  O_{\bf 20'}(x_2,Y_2) O_{\bf 20'}(x_3,Y_3) O_{\bf 20'}(x_4,Y_4)\rangle \,.
\end{align}
It can be decomposed into six terms, corresponding to the irreducible components in the tensor product of two $SU(4)$ representations
${\bf 20'} \times {\bf 20'}= {\bf 1} + {\bf 15}+{\bf 20'}+{\bf 84}+{\bf 105}+{\bf 175}$
\begin{equation}\label{G4-sum}
G_4 = 
{(N^2-1)^2\over 4(4\pi^2)^4}{ (y_{12}^2 y_{34}^2)^2 \over (x_{12}^2 x_{34}^2)^2}\sum_{{\cal R} \in {\bf 20'} \times {\bf 20'}}  G_{{\cal R}}(u,v)\,,
\end{equation}
where $x_{ij}^2=(x_i-x_j)^2$. Harmonic variables $y_{ij}^2=\epsilon_{ABCD} Y_i^{AB} Y_j^{CD}$ keep track of the $R-$charge dependence of the correlator
while the $x-$dependent prefactor carries the conformal weight of the operators.
Here $G_{{\cal R}}$ describes the contribution of all operators in the OPE of $ O_{\bf 20'}(1)  O_{\bf 20'}(2)$ that have $R-$charge corresponding to the $SU(4)$ representation $\cal R$. It depends on the cross ratios
\begin{align}
u=\frac{x_{12}^2x_{34}^2}{x_{13}^2 x_{24}^2}\,,\qqqquad v=\frac{x_{14}^2x_{23}^2}{x_{13}^2 x_{24}^2}\,,
\end{align}
as well as harmonic variables.  We do not display the $Y-$dependence for simplicity. 

The quadrilinear operators ${\cal A}_i$, defined in \re{quadrilinear}, are $SU(4)$ singlets and, therefore, they contribute to $G_{{\cal R}}(u,v)$
with ${\cal R} = {\bf 1}$. The general expression for $G_{{\bf 1}}(u,v)$ in $\mathcal N=4$ SYM is
\begin{align}\label{G1}
G_{{\bf 1}}(u,v) =  1 +\frac{2 u (v+1)}{3 \left(N^2-1\right) v}+ \frac{u^2 \left(3 N^2 \left(v^2+1\right)-3 v^2+4 v-3\right)}{60 \left(N^2-1\right) v^2} 
\\
 + \frac{u^2-8 u (v+1)+10 \left(v^2+4 v+1\right)}{60 (N^2-1) v^2} A(u,v)\,, \nonumber
\end{align}
where the contribution from the identity operator is exactly $1$ due to our choice of the normalization factor in \re{G4-sum}. The first line
on the right-hand side of \re{G1} describes the Born level contribution whereas the function $A(u,v)$ encodes all quantum corrections, both perturbative and non-perturbative. To leading order in both we have \cite{GonzalezRey:1998tk,Bianchi:1999ge}
\begin{equation}\label{A-fun}
A(u,v) = -2 {\coupling}  u v \bar D_{1111}(u,v)+ q \,Q u^2 v^2 \bar D_{4444}(u,v) + \dots \,, %
\end{equation}
where $\coupling=g^2 N/(4\pi^2)$ is the 't Hooft coupling constant,  $q=\e^{2\pi i\tau}$ is the instanton induced expansion parameter and~\footnote{This differs from the instanton correction considered in \cite{Arutyunov:2000im} by an overall factor of $(4\pi)^3$.}
\begin{align}\label{Qvalue}
Q = \frac{120}{\sqrt{\pi}(N^2-1)}\frac{\Gamma\left(N-\frac{1}{2} \right)}{\Gamma(N-1)}\,.
\end{align}
The dots on the right-hand side of \re{A-fun} denote subleading corrections suppressed by powers of $\coupling$ and $q$. The nontrivial
$u$ and $v$ dependence is described by the $\bar D$-functions 
\begin{align}
\bar D_{\Delta\Delta\Delta\Delta}(u,v)= \int_{-i \infty}^{i \infty} {dj_1 dj_2\over (2\pi i)^2} u^{j_1} v^{j_2} \Gamma^2(-j_1)\Gamma^2(-j_2)\Gamma^2(j_1+j_2+\Delta)\,.
\end{align}
The anti-instanton contribution to \re{A-fun} is given by the complex conjugated expression that we do not display for simplicity. 

Let us now consider the conformal partial wave expansion of the singlet channel contribution to \re{G4-sum}
\begin{equation}\label{cpw}
G_{{\bf 1}}(u,v)  = \sum_{\Delta,\ell} c^2_{\Delta,\ell} \, u^{\frac{\Delta-\ell}{2}} g_{\Delta,\ell}(u,v)\,,
\end{equation}
where the sum runs over conformal primaries with scaling dimension $\Delta$ and Lorentz spin $\ell$
transforming in the  singlet of $SU(4)$. The contribution of each conformal primary is given by the product of the square of the structure constant
$c_{\Delta,\ell}$ and the conformal block
\begin{equation}
\label{gdef}
g_{\Delta,\ell}(u,v) = \left(-\frac{1}{2} \right)^\ell \frac{1}{z-\bar z}\left[z^{\ell+1} k_{\Delta+\ell}(z) k_{\Delta-\ell-2} (\bar z) - \bar z^{\ell+1} k_{\Delta+\ell}(\bar z) k_{\Delta-\ell-2}(z)\right]\,,
\end{equation}
where $k_\beta(z) = {}_2F_1(\beta/2,\beta/2,\beta;z)$ and the auxiliary $z, \bar z$ variables are defined as $u=z\bar z$ and $v=(1-z)(1-\bar z)$. 

Matching \re{G1} and \re{cpw} we can determine the structure constants and the scaling dimensions of the conformal primary operators.
The prefactor  $u^{(\Delta-\ell)/2}$ on the right-hand side of \re{cpw} indicates that the small $u$ behaviour is controlled by the twist of the operator, $\tau=\Delta-\ell$. As follows from \re{G1} and \re{A-fun}, the leading instanton correction to  $G_{{\bf 1}}(u,v)$ scales at small $u$ as
$u^2 \bar D_{4444}(u,v) =O(u^2)$. This implies that the operators of twist two do not receive instanton corrections at $O(q)$ order \cite{Bianchi:1999ge}. Indeed, as was shown in \cite{Alday:2016tll}, the leading instanton correction to twist-two
operators of spin two scales as $O(g^2 q)$ whereas for higher spin it is suppressed at least by the power $g^2$.

In this paper, we are interested in the contribution to \re{cpw} from intermediate operators of twist four and spin zero that we shall denote as
$\Sigma_I$. Their scaling dimension
 takes the form  $\Delta_I = 4 +\gamma_I$ where the index $I$ enumerates the operators (which are degenerate in the free theory) and the anomalous dimensions $\gamma_I$ depend 
 on the two expansion parameters $a$ and $q$. At small $u$, the contribution of these operators to \re{cpw} scales as
 $u^{2+\gamma_I/2} =u^2 (1 + \frac12 \gamma_I \ln u + \dots)$. To determine the structure constants $c_I$ and anomalous dimensions 
 $\gamma_I$, we substitute \re{G1} into \re{cpw} and match term by term in a small $u,1-v$ expansion. Twist-four spin zero operators contribute at order  $u^2 (1-v)^0$ and $u^2 \ln u (1-v)^0$ on both sides of \re{cpw}. Contributions from descendants of twist-two operators, which have the same form, are automatically taken into account by the conformal blocks of the corresponding primaries. In this way, we obtain 
\begin{eqnarray}\nonumber
\label{OPEresults}
\sum_{I} c_{I}^2  &=& \frac{3 N^2-1}{30 \left(N^2-1\right)} + \dots \,, \\
\sum_{I} c_{I}^2 \gamma_I &=& -\frac{1}{N^2-1}\left( \frac25 \coupling + {18\over 35} q\, Q\right)  +\dots \,,  
\end{eqnarray}
where the sum runs over conformal primary operators of twist four and spin zero. Here the dots denote terms suppressed by powers of $a$ and $q$. 
To verify the relations \re{OPEresults}, it is sufficient to know $c_I$  and $\gamma_I$ at the lowest order in $a$ and $q$.
In what follows, we shall compute both quantities and demonstrate the validity of \re{OPEresults}.

As already mentioned in the introduction, the instanton contribution to four-point functions involving the Konishi operator vanishes at the semi-classical level. As we will see, this result is also consistent with our expressions for correlation functions \re{func}, but in a rather non-trivial way. 

\section{Instanton corrections to scalar operators}\label{instresults}

To define the conformal primary operators of twist-four and spin zero, $\Sigma_I$, we examine the two-point correlation function of 
quadrilinear operators ${\cal A}_i$ defined in (\ref{quadrilinear}). To leading order in $a=g^2 N/(4\pi^2)$ and $q=\e^{2\pi i\tau}$, it
has the following general form
\begin{equation}\label{AA}
\langle {\cal A}_i(x) {\cal A}_j(0) \rangle = \frac{16}{(4\pi^2)^4(x^2)^4}\left( H^{(0)}_{ij} -\coupling  H^{(1)}_{ij}  \log x^2 - q H^{\rm (inst)}_{ij} \log x^2 +\cdots \right)\,.
\end{equation}
The first two terms inside brackets describe the one-loop correction to the correlation function. They
were computed in \cite{Arutyunov:2002rs}  and the explicit expressions for the matrices
$H^{(0)}_{ij}$ and $H^{(1)}_{ij}$ can be found in appendix~\ref{appA}.
Later in this section we compute the
leading instanton correction $H^{\rm (inst)}_{ij}$ and compare it with analogous expression found in \cite{Kovacs:2003rt}. 

The twist-four conformal primary operators $\Sigma_I$ are given by a linear combination of the quadrilinear operators ${\cal A}_i$ 
and
satisfy the defining relation
\footnote{The spectrum of twist four scalar operators in the singlet representation of $SU(4)$ also includes operators build up from gauginos and the field strength. They will not, however, mix with the quadrilinear operators ${\cal A}_i$ at the order we are considering. } 
\begin{align}\label{Sigma-Sigma}
\vev{\Sigma_I(x) \Sigma_J(0)} = \delta_{IJ} {16 (N^2-1)^2 \over (4\pi^2 x^2)^4} \left(1- \gamma_I \log x^2 + \dots \right)\,.
\end{align}
The  anomalous dimensions $\gamma_I$ are  
given at leading order in $a$ and $q$ by the eigenvalues of the mixing matrix 
\begin{align}\label{Gamma}
\Gamma = (H^{(0)})^{-1}\left[\coupling H^{(1)} + q H^{\rm (inst)} \right]\,.
\end{align}
The corresponding eigenstates define the coefficients of the expansion of $\Sigma_I$ in the basis of ${\cal A}_i$ (see \re{below} below). 
 
\subsection{Results for $SU(2)$}

We start by computing the leading instanton correction to \re{AA} for the $SU(2)$ gauge group. Applying \re{instsc}, we have to evaluate 
the product of operators ${\cal A}_i(x) {\cal A}_j(0)$  in the instanton background and, then, integrate it over the collective coordinates
of instantons with the measure \re{measure}. 

The operators \re{quadrilinear} are built from scalars fields. For the one-instanton configuration in ${\cal N}=4$ SYM, these fields 
take the form \re{pro} with the leading term given for the $SU(2)$ gauge group by 
%
 \begin{equation}\label{phi2}               
\phi^{AB,(2)}_{ij}(x) = \frac{f(x)}{2 \sqrt{2}} \zeta_i^{[A} \zeta_j^{B]}\,,
\end{equation}
where $\zeta_\alpha^A(x)=\xi_\alpha^A+x_{\alpha \dot \alpha} \bar \eta^{\dot \alpha A}$ is a specific $x-$dependent linear combination of the fermion zero modes and the instanton profile $f(x)$ is  
\begin{equation}
f(x)=\frac{16 \rho^2}{((x-x_0)^2+\rho^2)^2}\,.
\end{equation}
The field \re{phi2} carries the $SU(2)$ indices $i,j=1,2$ and the $SU(4)$ indices $A,B=1,\dots 4$.  

We start by considering the one-instanton profile of the half-BPS and the Konishi operators, $O_{\bf 20'}(x,Y)$ and $K(x)$, defined in \re{bil}.  
These operators admit the expansion \re{K-exp} with the leading term given by \cite{Alday:2016tll,Alday:2016jeo}
\begin{align}
&O_{\bf 20'}(x,Y) = \frac{1}{g^2} \frac{128 \rho^4}{[\rho^2+(x-x_0)^2]^4} Y_{AB} Y_{CD} \left( \zeta^2 \right)^{AC}\left( \zeta^2 \right)^{BD}, \cr 
&K(x) = -\frac{1}{g^2} 3^2 \times 2^{15} \times \frac{\rho^6}{[\rho^2+(x-x_0)^2]^6} [\zeta(x)]^8+\dots \,,
\end{align} 
where $\left( \zeta^2 \right)^{AC}=\zeta^{\alpha A} \zeta^C_\alpha$ and $[\zeta(x)]^8=\prod_{A,\alpha} \zeta_\alpha^A$.

Let us now consider the operators \re{quadrilinear} on the instanton background. For 
the particular case of the $SU(2)$ gauge group the resulting expressions for ${\cal A}_i$ are not linearly independent
\begin{align}\label{rels}
{\cal A}_3 = {\cal A}_1-\frac12 {\cal A}_2\,,\qqqquad {\cal A}_4=\frac12 {\cal A}_2\,.
\end{align}
Since the expansion of a single scalar field \re{pro} is 
at least quadratic in fermion modes, the quadrilinear operators will have expansions starting at order eight,
${\cal A}_i = {\cal A}_i^{(8)} +  {\cal A}_i^{(12)} +  {\cal A}_i^{(16)}$. To compute the correlation function $\langle {\cal A}_i(1){\cal A}_j(2) \rangle_{\rm inst}$ using \re{instsc}, we have to retain only
terms containing $16$ fermion modes in the product of two operators
\begin{align}\label{AiAj}
\langle {\cal A}_i(x_1){\cal A}_j(x_2) \rangle_{\rm inst} &= \int d\mu_{\rm phys} {\cal A}_i^{(8)}(x_1){\cal A}_j^{(8)}(x_2)  \,.
\end{align}
For the operator ${\cal A}_1$ we have
 \begin{equation}\label{A1-8}
 {\cal A}_1^{(8)} = \frac{1}{g^4} \tr(\phi^{AB,(2)} \bar\phi_{AB}^{(2)} )  \tr( \phi^{CD,(2)} \bar\phi_{CD}^{(2)})\,,
\end{equation}
where $\bar \phi_{AB}^{(2)}= \frac{1}{2}\epsilon_{ABCD} \phi^{(2), CD}$. For the operator ${\cal A}_2$ we find using
\re{bil} and \re{quadrilinear} that ${\cal A}_2=K^2$. Then, it follows from \re{K-exp} that ${\cal A}_2^{(8)} =0$. As a result,
for the $SU(2)$ case only $\langle {\cal A}_1(1){\cal A}_1(2) \rangle_{\rm inst}$ is different from zero. 

Substituting \re{A1-8} into \re{AiAj} and performing the integration over the Grassmann variables we obtain
\begin{align}\label{int-div}
\langle {\cal A}_1(x_1){\cal A}_1(x_2) \rangle_{\rm inst} & = \frac{2^2 \times 3^4 \times 5^2}{\pi^{10}} \e^{2\pi i \tau} \int d^4 x_0 \int \frac{d\rho}{\rho^5} \frac{(x_{12}^2)^4 \rho^{16}}{(\rho^2+x_{10}^2)^8(\rho^2+x_{20}^2)^8}\,.
\end{align}
As expected the integral over bosonic collective coordinates develops a logarithmic divergence from the integration region $\rho \sim x_{i0}^2 \sim 0$. This signals that ${\cal A}_1$ acquires an anomalous dimension at order $O(q)$. To evaluate the integral \re{int-div} we have to introduce 
a regularization. To this end we modify the integration measure over the center of the instanton,
$\int d^4 x_0 \to \int d^{4-2\epsilon} x_0$. From \re{int-div}, we use the relation \re{2ptinst} to obtain 
\begin{align}\notag\label{A1-log}
\langle {\cal A}_1(1){\cal A}_1(2) \rangle_{\rm inst}  {}& =  - \frac{1350}{7\pi^8} (x_{12}^2)^{-4-\epsilon}\e^{2\pi i \tau}\Big(\frac{1}{\epsilon} + O(\epsilon^0)\Big)
\\
{}& 
= \frac{\log x_{12}^2}{(x_{12}^2)^4}  \frac{1350}{7\pi^8}\e^{2\pi i \tau}  + \cdots\,,
\end{align}
where in the second relation we retained only the term containing $\log x_{12}^2$. It is this term that contributes to the matrix $H^{\rm (inst)}$
in \re{AA}.

We would like to emphasize that the above mentioned regularization is different from the conventional dimensional regularization.
To implement the latter, one should start with $\mathcal N=4$ SYM in $D=4-2\epsilon$ dimensions and construct 
the instanton solution depending on $\epsilon$. This proves to be a nontrivial task given the fact that conformal symmetry of the theory
is broken for $\epsilon\neq 0$. One may wonder however whether the coefficient in front of $\log x_{12}^2$ in \re{A1-log} depends on 
the choice of regularization. To show universality of this coefficient, we can apply the dilatation operator $\mathbb D= x_1\partial_{x_1} + x_2\partial_{x_2} +8$ to the 
right-hand side of \re{int-div}. The resulting integral is finite and it yields the coefficient in front of $\log x_{12}^2$ in \re{A1-log}. 

Finally, we combine together the relations \re{A1-log} and \re{rels}, match them into \re{AA} and identify the mixing matrix defining the
leading instanton correction to $\langle {\cal A}_i(x) {\cal A}_j(0) \rangle$ for the $SU(2)$ gauge group
\begin{equation}
\label{instH}
H^{\rm (inst)}_{\rm SU(2)}=- \kappa_2 \left[ \scriptstyle
\begin{array}{cccc}
 1 & 0 & 1 & 0 \\
 0 & 0 & 0 & 0 \\
 1 & 0 & 1 & 0 \\
 0 & 0 & 0 & 0 \\
\end{array}
\right],
\end{equation} 
with $\kappa_2= 21600/7$. Before proceeding, let us make the following remark. The same mixing matrix was also computed in \cite{Kovacs:2003rt}. Our expression \re{instH} differs from the one presented there.~%
\footnote{The expression for $H^{\rm (inst)}$ found in \cite{Kovacs:2003rt} is not consistent with the linear relations \re{rels} for the $SU(2)$ gauge group. Furthermore, as we show below, the instaton corrections defined by \re{instH} are consistent with the OPE.}
 
The same analysis can be carried out for the three-point functions $\langle K(1) K(2) {\cal A}_i(3) \rangle$ and $\langle O_{\bf 20'}(1) O_{\bf 20'}(2) {\cal A}_i(3) \rangle$.  In the first case, the correlation function vanishes in the semiclassical approximation since the product of three operators
has $8\times 3$ fermion modes at least and gives zero upon integration over fermion modes,
\begin{align}
\langle K(1) K(2) {\cal A}_i(3) \rangle_{\rm inst} = 0\times  \e^{2\pi i \tau} \,.
\end{align}
In the second case, the calculation runs along the same lines as before. We replace operators by their expressions on the instanton background,
Eqs.~\re{K-exp} and \re{A1-8}, and integrate them over the collective coordinates with the measure \re{measure}
to obtain 
\begin{align}\notag
\label{inst3pt}
{}&
\langle O_{\bf 20'}(1) O_{\bf 20'}(2) {\cal A}_1(3) \rangle_{\rm inst} = (y_{12}^2)^2 \frac{3^4 \times 5}{\pi^{10}} \e^{2\pi i \tau}
\\
{}& \qqqquad \times
 \int d^4 x_0 \int \frac{d\rho}{\rho^5} \frac{(x_{13}^2 x_{23}^2)^2\,\rho^{16}}{(\rho^2+x_{10}^2)^4(\rho^2+x_{20}^2)^4(\rho^2+x_{30}^2)^8}\,.
\end{align}
For the operator ${\cal A}_2$ the same correlation function vanishes in the semiclassical approximation, 
\begin{align}\label{3-zero}
\langle O_{\bf 20'}(1) O_{\bf 20'}(2) {\cal A}_2(3) \rangle_{\rm inst}=0\times \e^{2\pi i \tau}\,.
\end{align}
For the operators
${\cal A}_3$ and ${\cal A}_4$ the answer is a linear combination of \re{inst3pt} and \re{3-zero}, by virtue of \re{rels}.
The integral \re{inst3pt} is divergent and needs to be regularized. As before, we do it by modifying the integration measure over $x_0$
\begin{align}\label{inst3pt1}
\langle O_{\bf 20'}(1) O_{\bf 20'}(2) {\cal A}_1(3) \rangle_{\rm inst} = - \frac{ 135}{28\pi^{8}} \e^{2\pi i \tau}
{(y_{12}^2)^2\over (x_{13}^2x_{23}^2)^2}\lr{{1\over \epsilon} + O(\epsilon^0)}
\end{align}
The details of the calculation can be found in appendix \ref{appB}. We also show in appendix \ref{appC} that (\ref{inst3pt}) leads to a finite contribution to the relevant OPE coefficients. 

\subsection{Generalisation to $SU(N)$}

So far our results are only valid for the $SU(2)$ gauge group. As explained in detail in \cite{Dorey:1998xe,Dorey:2002ik}, to leading order in the instanton expansion it is straightforward to generalise these results to the gauge group $SU(N)$. In this case the instanton has $8N$ fermion modes. These modes split into $16$ exact modes, whose contribution is identical to the one for the $SU(2)$ case, plus $8(N-2)$ non-exact modes. The contribution from the non-exact modes  factorises into a $N-$dependent factor. More precisely
\begin{equation}
\langle O(1) \cdots O(n)\rangle_{\rm 1-inst,\,SU(N)} = \frac{\kappa_N}{\kappa_2} \langle O(1) \cdots O(n)\rangle_{\rm 1-inst,\,SU(2)} 
\label{su2tosun}\,,
\end{equation}
where 
\begin{equation}\label{kappa}
\kappa_N = \frac{3 \Gamma^2(6)\Gamma(N-\frac{1}{2})}{7 \sqrt{\pi} \Gamma(N-1)} = {360 \over 7} (N^2-1) Q\,.
\end{equation}
In particular, applying  \re{su2tosun} to the relation \re{A1-log}, we find that the mixing matrix for the quadrilinear operators $\mathcal A_i$
at the instanton level is  given by
\begin{equation}
\label{instH1}
H_{\rm SU(N)}^{\rm (inst)} = { \kappa_N \over \kappa_2} H_{\rm SU(2)}^{\rm (inst)} \,,
\end{equation}
where $H_{\rm SU(2)}^{\rm (inst)}$ is given in \re{instH}.

\section{Consistency with higher point correlators}
\label{comparison}

In a generic CFT the operator product expansion allows us to write higher point correlation functions in terms of data appearing in lower order correlators. Given instanton corrections to two, three and four-point functions, a natural question is whether these results are consistent with the structure of the OPE. By performing a careful analysis we will answer this question affirmatively. 

\subsection{Solving the mixing problem}
\label{mixing}

In the previous section, we computed the leading instanton correction to the mixing matrix \re{Gamma}. Diagonalizing this matrix we can
construct the conformal primary operators $\Sigma_I$ and determine their anomalous dimensions $\gamma_I $ to leading order in $a$ and $q$
\begin{align}\label{below}
\Gamma_{ij} \,\psi_{i,I} =\gamma_I \psi_{i,I}\,,\qqqquad \Sigma_I(x) = \psi_{i,I} \mathcal A_i(x)\,,
\end{align}
where the index $I$ enumerates the eigenstates $\psi_{i,I}$. The normalization of the eigenstates $ \psi_{i,I}$ is fixed by relation \re{Sigma-Sigma}. 
It is straightforward to verify, with the help of \re{AA}, that $\Sigma_I$ defined in this way satisfies \re{Sigma-Sigma} provided that the 
eigenstates are normalized as
\begin{align}
 \psi_{i,I} H^{(0)}_{ij} \psi_{j,I} = \delta_{IJ} (N^2-1)^2\,.
\end{align}
The diagonalization of the mixing matrix \re{Gamma} is very cumbersome for general $N$. In the following we consider two separate cases. 
First the case $N=2$ and then the expansion in $1/N$ around large $N$.

\subsection*{$SU(2)$ gauge group}
In this case, the analysis is particularly simple since there are only two linearly independent operators, see \re{rels}. 
Substituting \re{instH}, \re{H0} and \re{H1} into \re{Gamma} we check that for $N=2$ the mixing matrix has rank $2$ indeed.  
Its eigenvalues are given at leading order by
\begin{eqnarray} 
\label{su2an}
\gamma_1 &=&  -\frac{3}{2} \coupling - \frac{q \kappa_2}{95}\,, \qqqquad
\gamma_2 = 8 \coupling - \frac{q \kappa_2}{1710}\,.
\end{eqnarray}
Higher order corrections to these relations, proportional to $q^2, \coupling^2, q \coupling$, etc, will not be relevant for our discussion. 
Applying the second relation in \re{below}, we find the explicit expressions for the operators $\Sigma_i$. They are given at leading 
order by
%
\begin{align}\nonumber
\label{eigenstatessu2}
\Sigma_1 & =  \mathcal N_1 \left[ 3 {\cal A}_1-{\cal A}_2  - {\kappa_2\over 570}  \frac{q}{\coupling} {\cal A}_1 \right]\,,
\\
\Sigma_2 & = \mathcal N_2 \left[ 2 {\cal A}_1-7 {\cal A}_2 -{4 \kappa_2 \over 1995}  \frac{q}{\coupling}{\cal A}_1 \right]\,,
\end{align}
with the normalization factors $1/\mathcal N_1^2= 16(95-{7 q \kappa_2}/{(57 a)})$ and $1/\mathcal N_2^2= 16({760}+{64 q \kappa_2}/{(399 a)})$.

Higher order corrections to \re{eigenstatessu2} are proportional to $q,\coupling,$ etc. Note a very important point. Instanton effects induce corrections
to the conformal operators \re{eigenstatessu2} (as well as to their OPE coefficients) that are proportional to $q/\coupling$. At the same time, such corrections are absent on the right-hand
side of the sum rules \re{OPEresults}. This implies that, for consistency with the operator algebra, $O(a/q)$ terms should cancel against each
other in the sum over conformal primary operators on the left-hand side of \re{OPEresults}. Furthermore, we will see that such corrections are actually crucial for consistency with the OPE. 

\subsection*{Large $N$}

In this case, the eigevalues of the mixing matrix \re{Gamma} are given by
\begin{eqnarray}\notag\label{anom}
\gamma_1 &=& \coupling\left( -\frac{10  }{N^2}+\cdots \right)  +q \left( - { \kappa_N\over 10N^4}  + \dots \right),
\\[2mm] \notag
\gamma_2 &=&  \coupling \,\big( 6+\cdots \big) +q \left( - \frac{\kappa_N}{2 N^6} + \dots \right), \\
\gamma_{\pm} &=& \coupling \left( \frac{13}{4} \pm \frac{\sqrt{41} }{4} + \cdots \right)  +q \left( - { \kappa_N\over N^4} \left( \frac{7}{120} \mp \frac{9}{40 \sqrt{41} }   \right) + \dots \right), 
\end{eqnarray}
where dots denote corrections suppressed by powers of $1/N^2$. Notice that the eigenvalues satisfy the following relation 
at weak coupling
\begin{align}
\gamma_1< \gamma_-< \gamma_+ < \gamma_2\,.
\end{align}
Viewed as eigenvalues of the dilatation operators, the functions $\gamma_i(a)$ cannot cross each other. This
implies that the same relation holds for an arbitrary coupling $a$. For large values of $a$ and $N$ it has been argued, see \cite{Arutyunov:2002rs}, that $\Sigma_2$ and $\Sigma_{\pm}$ acquire a large anomalous dimension, while $\Sigma_1$ is dual to a multiparticle supergravity state and has
a finite scaling dimension. 

As for the eigenstates of the mixing matrix, following a tedious but otherwise standard procedure we find from \re{below}
 %
%
%
\begin{align}\notag\label{eigenstates}
\Sigma_1 {}& = \mathcal N_1\left[-6 A_1+A_2 -{q \kappa_N\over 160\coupling N^4} {(59 A_3+6 A_4)}  \right],
\\[2mm]\notag
\Sigma_2 {}& = \mathcal N_2\left[A_2-{q \kappa_N\over 120\coupling N^5} {(-6 A_1+A_2-15 A_3+18 A_4)
 } \right],
\\[2mm]\notag
\Sigma_\pm {}& = \mathcal N_\pm\bigg[\frac{5\mp\sqrt{41}}{4}  A_3+A_4 -{43\mp 7\sqrt{41}\over 8N} A_1
\\
{}&
- \left(53\mp 9 \sqrt{41}\right){q \kappa_N\over  640\coupling N^4}
  \left(A_1-\frac{A_2}{6}+\frac{319\pm 59 \sqrt{41}}{492}
  A_3+\frac{206\pm 6 \sqrt{41}}{492}  A_4\right)   \bigg],
\end{align}
with normalization factors $1/\mathcal N_1^2 =5760$, $1/\mathcal N_2^2 =288$ and $1/\mathcal N_\pm^2 = 2(369\mp 51\sqrt{41})$.
Subleading corrections to \re{eigenstates} are suppressed by powers of $a$, $q$ and $1/N$. Higher powers in $1/N$ are not explicitly shown, since they are not particularly enlightening, but they are necessary for consistency with the OPE. 

As before, the instaton corrections to $\Sigma_i$ induce terms proportional to ${q}/{\coupling}$. They are crucial for consistency with the OPE. 
The anti-instanton corrections to \re{anom} and \re{eigenstates} are given by complex conjugated expressions with $q\to \bar q$.

\subsection{Consistency conditions}

Let us now perform the comparison with the results in section \ref{OPEanal}. In order to proceed, we compute the canonically normalised OPE coefficients between two half-BPS operators $O_{\bf 20'}$ and the conformal operators $\Sigma_I$
\begin{align}\label{c-gen}
c_I^2 = \frac{\langle O_{\bf 20'}O_{\bf 20'} \Sigma_I \rangle^2}{\langle O_{\bf 20'}O_{\bf 20'} \rangle^2 \langle \Sigma_I  \Sigma_I \rangle}\,,
\end{align}
where the dependence on the coordinates of operators is neglected. The two-point functions entering the denominator are given by
 \re{Sigma-Sigma} and \re{O-2pt}. To find the three-point function in the numerator, we use \re{below} to get
\begin{align}\label{LL}
\langle O_{\bf 20'}O_{\bf 20'} \Sigma_I \rangle = \psi_{i,I} \langle O_{\bf 20'}O_{\bf 20'} \mathcal A_i \rangle\,,
\end{align}
where the correlation functions on the right-hand side are given in the Born approximation by \re{3pt}.

Given the explicit form of the operators, Eqs.~\re{eigenstatessu2} and \re{eigenstates}, the OPE coefficients admit an expansion of the form
\begin{eqnarray}\label{c-exp}
c_{I}  = c^{(0)}_I + \frac{q}{\coupling} c^{\rm (inst)}_{I} + \cdots\,,
\end{eqnarray}
where $c^{(0)}_I$ is the Born level approximation and $c^{\rm (inst)}_{I}$ defines the leading instanton correction.
Here the dots denote corrections suppressed by powers of $q$ and $a$, they will not be relevant for our discussion. 
It is easy to see that the dependence of the OPE coefficients \re{c-gen} on $q/\coupling$ only comes from the expansion coefficients
$\psi_{i,I}$ in \re{LL}, the leading corrections to $\langle \Sigma_I  \Sigma_I \rangle$ and $ \langle O_{\bf 20'}O_{\bf 20'} \mathcal A_i \rangle$
are linear in $a$ and $q$. Therefore, computing the leading correction to \re{c-exp}, we are allowed to replace these
correlation functions by their Born level expressions, Eqs.~\re{Sigma-Sigma} and \re{3pt}.

 The OPE coefficients \re{c-gen} together with the anomalous dimensions $\gamma_I$ obtained in section \ref{mixing} lead to the following results: 

\subsection*{Gauge group $SU(2)$}
For the $SU(2)$ gauge group there are only two conformal operators \re{eigenstatessu2}. At leading order we obtain
\begin{eqnarray} \label{c12}\notag
c_{1}^2= \frac{20}{171} - \frac{4 \kappa_2}{308655} \frac{q}{\coupling}+ \cdots \,,  \\
c_{2}^2= \frac{1}{190} +\frac{4 \kappa_2}{308655} \frac{q}{\coupling} +\cdots \,.
\end{eqnarray}
Combining these results with the anomalous dimensions (\ref{su2an}) we obtain 
\begin{eqnarray} 
\label{c02}
\sum_{I=1,2} c_{I}^2  &=& \frac{11}{90}\,, \qqqquad 
\sum_{I=1.2} c_{I}^2 \gamma_I = -\frac{2}{15} \coupling - q \frac{\kappa_2}{900}\,.
\end{eqnarray}
%

\subsection*{Large $N$}
The same procedure can be carried out for the large $N$ expansion. In this case we obtain from \re{anom} and 
\re{eigenstates}
\begin{eqnarray}\nonumber
\label{c0n}
\sum_{I} c_{I}^2  &=& \frac{1}{10} +\frac{1}{15 N^2}+\frac{1}{15 N^4} + \cdots \,,\\
\sum_{I} c_{I}^2 \gamma_I &=&\coupling \left( -\frac{2}{5 N^2} -\frac{2}{5 N^4} + \cdots \right) -q \kappa_N \left( \frac{1}{100 N^4}+ \frac{1}{50N^6} + \cdots \right).
\end{eqnarray}


The relations (\ref{c02}) and (\ref{c0n}) have to be compared to (\ref{OPEresults}). Using the explicit expression for $\kappa_N$ given in (\ref{kappa}),  we observe a perfect agreement in both cases, for $N=2$ and at large $N$. We would like to stress that the corrections to the eigenstates of the form $q/a$ are crucial in these comparisons. Finally, since the OPE coefficients and the anomalous
dimensions on the left-hand side of the last two relations were found from two- and three-point correlation functions, this result represents a nontrivial consistency check of the
approach to computing instanton corrections that we employed in this paper.


\subsection{Four-point correlators involving the Konishi operator}

We can perform a similar analysis for four-point correlation functions involving the Konishi operator, $\langle K K K K \rangle$ and $\langle O_{\bf 20}  O_{\bf 20} K K \rangle$. The main difference with the four-point correlation function of half-BPS operators \re{G4} is that they do not receive  instanton corrections at the 
leading order $O(q)$. Indeed, as follows from \re{K-exp}, the product of Konishi and half-BPS operators contains more than $16$
fermion modes and vanishes upon integration in \re{instsc}. On the other hand, the twist-four conformal operators $\Sigma_I$ arise in the OPE
of Konishi operators, 
\begin{align}
K(x) K(0) \sim \sum_I k_I \,\Sigma_I(0) + \dots
\end{align}
and, therefore, contribute to the above mentioned correlation functions. 

The OPE coefficients $k_I$ receive instanton corrections and have the same general form as \re{c-exp}. 
Then, the vanishing of the leading instanton corrections to $\langle K K K K \rangle$ and $\langle O_{\bf 20}  O_{\bf 20} K K \rangle$ implies that the
$O(q/a)$ terms should cancel in the  following combinations 
\begin{align}\label{sums}
\sum_I k_I^2 \,,\qqquad \sum_I k_I^2 \gamma_I/a \,,\qqquad \sum_I k_I c_I \,,\qqquad \sum_I k_I c_I \gamma_I/a \,.
\end{align}
The structure constants $c_I$ and the anomalous dimensions $\gamma_I$ were computed in the beginning of this section. 
Using the results of appendix \ref{appA} (see Eq.~\re{3pt}) we can compute the OPE coefficients 
$k_I\sim \vev{KK\Sigma_I}=\sum \psi_{i,I} \vev{KK\mathcal A_i}$
to leading order in $q/a$.

For instance, for the case of $SU(2)$ gauge group we use \re{eigenstatessu2}, \re{3pt} and \re{O-2pt} to find 
\begin{eqnarray}\notag
k_{1}^2&=& \frac{5}{684} - \frac{ \kappa_2}{61731} \frac{q}{\coupling}+ \cdots \,, \\
k_{2}^2&=& \frac{5}{38} + \frac{ \kappa_2}{61731} \frac{q}{\coupling} +\cdots \,.
\end{eqnarray}
Combining these relations together with \re{su2an} and \re{c12}, we verify that the leading instanton corrections
disappear in all four sums \re{sums}. This is a rather non-trivial result since each individual term in the sums does depend on $q/a$. 

Thus, we conclude that the approach followed in this paper leads to expressions for the OPE coefficients and the anomalous
dimensions of quadrilinear operators \re{quadrilinear} that are fully consistent with the structure of the OPE.

\section{Conclusions}\label{concl}

In this paper we have revisited the computation of instanton effects to various correlation functions in ${\cal N}=4$ SYM
and resolved a controversy existing in the literature regarding their consistency with the OPE and conformal symmetry. 

Since instantons preserve conformal invariance of  ${\cal N}=4$ SYM, the obtained expressions for instanton corrections to correlation functions should be consistent with conformal symmetry. To check this property, we examined the conformal partial wave decomposition of four-point correlators involving combinations of the half-BPS operator $O_{\bf 20'}$ and the Konishi operator $K$ and isolated the contribution from the conformal primary operators
built from the twist-four quadrilinear operators \re{quadrilinear}. We demonstrated that the leading instanton correction to this contribution is indeed consistent 
with the conformal symmetry and computed the corresponding corrections to the OPE coefficients and the scaling dimensions of the quadrilinear operators.
Although the later corrections are perfectly finite, their computation involves divergent integrals over the collective coordinates of instantons which need to be regularised. We do this by dimensionally regularizing the integral over the position of the instanton. Our computation shows that this regularization procedure yields expressions for the OPE coefficients and anomalous dimensions which are in perfect agreement with conformal symmetry.  

There are several directions which can be pursued. The spectrum of the dilatation operator in $\mathcal N=4$ SYM is believed to be
invariant under modular $S-$duality transformations. At weak coupling, the Konishi operator $K$ and quadrilinear operators $\Sigma_I$ are the lowest eigenstates of this operator in the $SU(4)$ singlet sector. One of the byproducts of our analysis is the determination of the leading instanton correction to the scaling dimension of the later operators, $\Delta_{\Sigma_I} =4+\gamma_I(q,\bar q)$ with $\gamma_I$ given by \re{su2an} and \re{anom}. For the 
Konishi operator, $\Delta_K = 2+ \gamma_K(q,\bar q)$, the analogous correction has been computed in \cite{Alday:2016tll,Alday:2016jeo}. It would be interesting to understand the modular properties of  $\Delta_{\Sigma_I}(q,\bar q)$ and $\Delta_K(q,\bar q)$. $S-$duality suggests that they should be modular invariant functions.
A related question is that of level crossing (or avoidance) of the first two levels of the dilatation operator of the theory, namely $\Delta_K(q,\bar q)$ and $\Delta_{\Sigma_1}(q,\bar q)$. We expect instanton corrections to play a fundamental role in whatever mechanism is at play.~\footnote{Although there has been some progress understanding this issue for large $R-$charges  \cite{Minahan:2014usa}, at large $N$  \cite{Korchemsky:2015cyx} and for operators with large spin  \cite{Korchemsky:2015cyx,Alday:2013cwa}, the general case with no large parameters remains to be understood.}  

It would be also interesting to compute subleading $O(aq)$ instanton corrections to four-point correlation functions \re{G4}. Although this would require taking into account quantum fluctuations around the instanton background, some pieces of the answer can be deduced from our analysis. For instance, in a small $u$ expansion, the terms of the form $O(u^2 \log^2 u)$ are fixed by the conformal symmetry up to the factor   $\sum_I c_{I}^2 \gamma_I^2$, which can be exactly computed from the results in this paper.  

\section*{Acknowledgements}  

The work of L.F.A. was supported by ERC STG grant 306260. L.F.A. is a Wolfson Royal Society Research Merit Award holder.  This work of G.P.K. was supported in part by
the French National Agency for Research (ANR) under contract StrongInt (BLANC-SIMI-
4-2011).

\appendix
\section{Correlation functions in the Born approximation}
\label{appA}
Free propagators of scalar fields $\phi^{AB} = \phi_a^{AB} T^a$ are given by
\begin{eqnarray}\label{free}
\langle \phi^{AB}_a(x_1) \phi^{CD}_b(x_2)\rangle = g^2 D(x_{12}) \epsilon^{ABCD} \delta_{ab}\,,
\end{eqnarray}
where $a,b=1,\cdots N^2-1$ denote color indices in the adjoint representation of the $SU(N)$ gauge group  and we have introduced
$D(x) = 1/(4\pi^2 x^2)$ and $x_{12}=x_1-x_2$. Propagators involving conjugated scalar fields $\bar \phi_{AB}= \frac{1}{2} \epsilon_{ABCD} \phi^{CD}$ can be deduced from the one above. The $SU(N)$ generators are normalised as 
\begin{equation}
\tr \big( T^a T^b \big) = \frac{1}{2} \delta^{ab},\qqqquad \tr 1 = N\,.
\end{equation}  
Traces involving a higher number of generators can be simplified using the identities
\begin{eqnarray}\notag
&& \tr \left(T^a A \right) \tr\left(T^a B \right) =\frac{1}{2} \tr \left( A B \right) - \frac{1}{2N} \tr A \tr B \,,\\
&& \tr \left(T^a A T^a B \right) =\frac{1}{2}\tr A \tr B  - \frac{1}{2N} \tr \left( A B \right)\,.
\end{eqnarray}
We now present the results for various correlators in the Born approximation. 
For the bilinear scalar operators \re{bil} we have
\begin{eqnarray}\notag\label{O-2pt}
&& \langle O_{\bf 20'}(1) O_{\bf 20'}(2) \rangle = \frac12(N^2-1) (y_{12}^2)^2 D^2(x_{12})\,, \\
&& \langle K(1) K(2) \rangle = 12(N^2-1)D^2(x_{12}) \,,
\end{eqnarray}
where  $y_{12}^2 = \epsilon^{ABCD}Y_{1,AB} Y_{2,CD}$.

Next we consider three-point correlators between two bilinear operators  \re{bil} and quadrilinear scalar operators \re{quadrilinear}. 
For correlators involving two half-BPS operators we obtain in the Born approximation
\begin{align}\notag\label{3pt}
&  \langle O_{\bf 20'}(1)O_{\bf 20'}(2) {\cal A}_{i}(3) \rangle =  8 a_i (N^2-1) (y_{12}^2)^2D^2(x_{13})D^2(x_{23}),
\\[2mm]
&  \langle K(1) K(2) {\cal A}_{i}(3) \rangle = 48 b_i (N^2-1)  D^2(x_{13})D^2(x_{23})\,,
\end{align}
with the coefficients $a_i$ and $b_i$ given by
\begin{align}\notag
& a_1 = N^2\,,&&  a_2= 2\,, && a_3= 2(N^2-1)/N\,,&& a_4=(N^2-2)/N\,,
\\[2mm]
& b_1=N^2+6\,,&& b_2=2(3N^2-2)\,,&& b_3=2(N^2-4)/N\,,&& b_4=(7N^2-8)/N\,.
\end{align}
%
%
%
Finally, the two-point correlation functions of the operators \re{quadrilinear} are given perturbatively by~\cite{Arutyunov:2002rs}
\begin{equation}\label{A-A}
\langle {\cal A}_{i}(1){\cal A}_{j}(2) \rangle =16  D^4(x_{12}) \left[H_{ij}^{(0)}  -\coupling H_{ij}^{(1)} + O(\coupling^2)\right]\,,
\end{equation}
where $\coupling = g^2 N/(4\pi^2)$ is the 't Hooft coupling constant and the matrix $H_{ij}^{(0)}$ takes the form
 \begin{equation}\label{H0}
H^{(0)}= 3(N^2-1)\left[
\begin{array}{cccc}
 \frac{1}{2} \left(7 N^2+2\right) & N^2+6 & \frac{7 N^2-8}{N} & \frac{9 N^2-16}{2 N} \\
 N^2+6 & 2 \left(3 N^2-2\right) & \frac{2 \left(N^2-4\right)}{N} & \frac{7 N^2-8}{N} \\
 \frac{7 N^2-8}{N} & \frac{2 \left(N^2-4\right)}{N} & \frac{3 N^4-8 N^2+24}{N^2} & \frac{N^4-16 N^2+48}{2 N^2} \\
 \frac{9 N^2-16}{2 N} & \frac{7 N^2-8}{N} & \frac{N^4-16 N^2+48}{2 N^2} & \frac{7 N^4-32 N^2+96}{4 N^2} \\
\end{array}
\right].
 \end{equation}
 For completeness, we also present one-loop correction to \re{A-A} found in \cite{Arutyunov:2002rs}
 \begin{equation}\label{H1}
H^{(1)}=-\frac{3}{2}(N^2-1) \left[
\begin{array}{cccc}
 13-2 N^2 & -6 \left(2 N^2+7\right) & \frac{21 N^2+16}{N} & \frac{32-53 N^2}{2N} \\
 -6 \left(2 N^2+7\right) & -12 \left(6 N^2+1\right) & \frac{6 \left(N^2+16\right)}{N} & -\frac{3 \left(33 N^2-32\right)}{N} \\
 \frac{21 N^2+16}{N} & \frac{6 \left(N^2+16\right)}{N} & \frac{-11 N^4+96 N^2-128}{N^2} & \frac{3 N^4+112 N^2-256}{2 N^2} \\
 \frac{32-53 N^2}{2N} & -\frac{3 \left(33 N^2-32\right)}{N} & \frac{3 N^4+112 N^2-256}{2 N^2} & \frac{-59 N^4+64 N^2-512}{4 N^2} \\
\end{array}
\right].
\end{equation}

\section{Dimensionally regularized integrals}\label{appB}

The instanton correction to two-point correlation function \re{int-div} involves the following integral
\begin{align}\notag
\label{2ptinst0}
I_\alpha(x_1,x_2) 
{}& =\int d^{4-2\epsilon} x_0 \int \frac{d\rho}{\rho^5} \frac{(x_{12}^2)^\alpha \rho^{2 \alpha}}{(\rho^2+x_{10}^2)^\alpha(\rho^2+x_{20}^2)^\alpha} 
\\
{}& =  -\frac{1}{\epsilon} (x_{12}^2)^{-\epsilon} \pi^{2-\epsilon} \frac{\Gamma(\alpha-2)\Gamma(\alpha+\epsilon)\Gamma^2(1-\epsilon)}{\Gamma^2(\alpha)\Gamma(1-2\epsilon)}\,,
\end{align}
evaluated for $\alpha=8$. Here in the first relation we dimensionally regularized the integral over the position of the instanton.
Divergences appear as poles in $\epsilon$ and come from integration over  the small size instantons $\rho\to 0$ located close to external
points, $x_{10}^2\to 0$ or $x_{20}^2\to 0$. Expanding \re{2ptinst0} around small $\epsilon$ we get for $\alpha=8$
\begin{align}\label{2ptinst}
I_8(x_1,x_2) =   {\pi^2\over 42}   
\lr{-{1\over\epsilon} + \ln x_{12}^2 + \dots}\,.
\end{align}
Let us consider the integral  (\ref{inst3pt}) defining the leading  instanton correction  to three-point functions 
\begin{equation}\label{I3}
I_8(x_1,x_2,x_3) = \int d^{4-2\epsilon} x_0 \int \frac{d\rho}{\rho^5} \frac{ (x_{13}^2 x_{23}^2)^{4}\rho^{16}}{(\rho^2+x_{10}^2)^{4}(\rho^2+x_{20}^2)^{4}(\rho^2+x_{30}^2)^8}\,.
\end{equation}
A simple power counting shows that for $\epsilon\to 0$ the integral develops a logarithmic divergence for $\rho\to 0$ and $x_{30}^2\to 0$.
As before, it appears as a pole in $\epsilon$. To find the residue at this pole, we examine the contribution to \re{I3} coming from the region
of small $\rho$ and $x_{30}^2$ that we denote as $\Omega$
\begin{align}
I_8(x_1,x_2,x_3) \sim  \int_\Omega d^{4-2\epsilon} x_{0'}  \frac{d\rho}{\rho^5}\frac{ \rho^{16}}{ (\rho^2+x_{0'}^2)^8}\,,
\end{align}
where $x_0'=x_{30}$. In the similar manner, the divergent contribution to \re{2ptinst0} coming from $\rho\to 0$ and $x_{10}^2\to 0$ or  
$x_{20}^2\to 0$ is given by
\begin{align}
I_8(x_1,x_2) \sim  2\int_\Omega d^{4-2\epsilon} x_{0'}  \frac{d\rho}{\rho^5}\frac{ \rho^{16}}{ (\rho^2+x_{0'}^2)^8}\,.
\end{align}
Comparing the last two relations we deduce that the divergences cancel in the difference of integrals 
$ 
I_8(x_1,x_2,x_3) - \frac12 I_8(x_1,x_2) = O(\epsilon^0) \,.
$
Together with \re{2ptinst} this immediately leads to
\begin{align}\label{I3-1}
I_8(x_1,x_2,x_3) =  -\frac{\pi^2}{84 \epsilon} + O(\epsilon^0)\,.
\end{align}
Substituting \re{I3} and \re{I3-1} into \re{inst3pt} we arrive at \re{inst3pt1}.
 
\section{Finiteness of instanton corrections to structure constants}
\label{appC}

According to \re{A1-log} and \re{inst3pt1}, the instanton contribution to the correlation functions contain ultraviolet divergences.
They appear as poles $1/\epsilon$ in the parameter of dimensional regularization and produce corrections to the scaling dimensions of unprotected operators. We verify in this appendix that ultraviolet divergences cancel in the expression for the OPE coefficients \re{c-gen}.

We start with the three-point functions $\langle O_{\bf 20'}O_{\bf 20'} \Sigma_I \rangle$. As follows from \re{LL}, it is given by a linear combination
of three-point functions of quadrilinear operators \re{quadrilinear}. To the lowest order in $a$ and $q$ the later functions have the following form
\begin{align}\label{OOA-gen}
\langle O_{\bf 20'}(x_1)O_{\bf 20'}(x_2) {\cal A}_{i}(0) \rangle =   {8  (N^2-1)(y_{12}^2)^2 \over (4\pi^2)^4 (x_1^2x_2^2)^2} \left[f_i^{(0)} + {1\over \epsilon}\lr{a f_i^{(1)}
+q f_i^{\rm (inst)}}+\dots  \right],
\end{align}
where the Born level contribution $f_i^{(0)}=  a_i $ is defined in \re{3pt}. The one-loop and instanton corrections are described by  $f_i^{(1)}$ and $f_i^{\rm (inst)}$, respectively. Substituting \re{OOA-gen} into \re{LL} we find that $\langle O_{\bf 20'}O_{\bf 20'} \Sigma_I \rangle$ can be factor out 
into the product of UV divergent and regular contributions
\begin{align}\label{Z-3}
\langle O_{\bf 20'}O_{\bf 20'} \Sigma_I \rangle ={\langle O_{\bf 20'}O_{\bf 20'} \Sigma_I \rangle}_{\! R} \left[ 1+ {1\over \epsilon}\lr{a  {(\psi,f^{(1)})\over (\psi,f^{(0)})} 
+q {(\psi,f^{\rm (inst)})\over (\psi,f^{(0)})}  }+\dots\right]\,,
\end{align}
where we use a shorthand notation for $(\psi,f) = \psi_i f_i$. 
For the two-point correlation function $\vev{\Sigma_I(x) \Sigma_I(0)}$ we find from \re{Sigma-Sigma} in the similar manner
\begin{align}\label{Z-2}
\vev{\Sigma_I \Sigma_I } = {\vev{\Sigma_I \Sigma_I }}_{\! R} \left(1+ {\gamma_I\over \epsilon}  + \dots \right)\,,
\end{align}
where we took into account that $\ln x^2$ term on the right-hand side of \re{Sigma-Sigma} originates from the small $\epsilon$ expansion of 
$(x^2)^{-\epsilon}/\epsilon$.

Combining together \re{Z-3}, \re{Z-2} and \re{c-gen} we find that the OPE coefficients $c_I$ remain finite for $\epsilon\to 0$ provided that
\begin{align}
a (\psi,f^{(1)}) 
+q  (\psi,f^{\rm (inst)}) = \frac12 \gamma_I(\psi,f^{(0)}) = \frac12 (\Gamma\psi,f^{(0)})\,,
\end{align}
where in the second relation we took into account \re{below}. Replacing the mixing matrix $\Gamma$ with its explicit expression \re{Gamma} and
matching the coefficients in front of $a$ and $q$, we arrive at 
\begin{align}\notag\label{ff}
{}& f_i^{(1)} = \frac12\left[ H^{(1)}  (H^{(0)})^{-1}\right]_{ij}  f_j^{(0)}\,,\\
{}& f_i^{\rm (inst)} =\frac12 \left[ H^{\rm (inst)}  (H^{(0)})^{-1}\right]_{ij}  f_j^{\rm (0)}\,.
\end{align} 
These relations fix the UV divergent part of the correlation function \re{OOA-gen}. 

Let us verify relations \re{ff} for the $SU(2)$ gauge group. In this case, we use the explicit expressions for the mixing matrices \re{instH} and \re{H0} \footnote{More precisely, since there are only two conformal operators, we need the upper $2\times 2$ blocks of these matrices.} together with
$f_j^{(0)} =(4,2)$ to get
\begin{align}
 f_1^{\rm (inst)} =  -{\kappa_2\over 60} \,,\qqqquad  f_2^{\rm (inst)} = 0\,.
\end{align}
Substituting these expressions into \re{OOA-gen} and putting $N=2$ we reproduce  \re{3-zero} and  \re{inst3pt1}.

\bibliographystyle{JHEP} 


\providecommand{\href}[2]{#2}\begingroup\raggedright\endgroup

\end{document}